\documentclass[12pt]{article} 

\usepackage{amsmath}
\usepackage{amssymb}
\usepackage[utopia]{mathdesign}
\usepackage{eucal}
 \usepackage{cite}
 \usepackage{hyperref}
\usepackage[height=8.8in,width=6.45in]{geometry}
\usepackage{tikz}

\numberwithin{equation}{section}

\newcommand{\cC}{\mathcal{C}}

\newcommand{\cH}{\mathcal{H}}

\newcommand{\cL}{\mathcal{L}}
\newcommand{\cM}{\mathcal{M}}
\newcommand{\cN}{\mathcal{N}}

\newcommand{\cZ}{\mathcal{Z}}

\newcommand{\bC}{\mathbb{C}}

\newcommand{\bR}{\mathbb{R}}

\newcommand{\bZ}{\mathbb{Z}}

\newcommand{\fg}{\mathfrak{g}}

\def\vev#1{\left\langle #1 \right\rangle}
\def\vvevv#1{\left\langle\left\langle #1 \right\rangle\right\rangle}

\def\U{\mathrm{U}}
\def\SU{\mathrm{SU}}

\def\SU{\mathrm{SU}}

\def\Tr{\mathrm{Tr}}
\def\rank{\mathrm{rank}}

\let\tilde\widetilde

\begin{document}

\begin{titlepage}

\begin{flushright}
IPMU-13-0163\\
UT-13-31\\
\end{flushright}

\vskip 3cm

\begin{center}
{\Large \bf
On the 6d origin of\\[1em]
discrete additional data of 4d gauge theories
}

\vskip 2.0cm
  Yuji Tachikawa
\vskip 1.0cm

\begin{tabular}{ll}
  & Department of Physics, Faculty of Science, \\
& University of Tokyo,  Bunkyo-ku, Tokyo 133-0022, Japan, and\\
  & Institute for the Physics and Mathematics of the Universe, \\
& University of Tokyo,  Kashiwa, Chiba 277-8583, Japan\\
\end{tabular}

\vskip 1cm

\textbf{Abstract}
\end{center}

Starting with a choice of gauge algebras, specification of a 4d gauge theory involves additional data, namely the gauge groups  and the discrete theta angles. Equivalently, one needs to specify the set of charges of allowed line operators. 
In this note, we study how these additional data are represented in 6d, when the 4d theory in question is an $\cN{=}4$ super Yang-Mills theory or an $\cN{=}2$ class $S$ theory. 
We will see that the $\mathbb{Z}_N$ symmetry of the so-called $T_N$ theory plays an important role.

As a byproduct, we will find that the superconformal index of class $S$ theories can be refined so that it can give 2d $q$-deformed Yang-Mills theory with different gauge groups associated to the same gauge algebra.

\medskip
\noindent

\end{titlepage}



\section{Introduction}

As is well-known,
specification of a gauge theory requires a choice of the gauge group and not just its Lie algebra.
This is particularly true when the spacetime geometry is nontrivial, or equivalently when general line operators are considered, which create holes in the spacetime. 
The choice of the gauge group does not yet quite fix the quantum field theory considered, and one needs to specify the set of allowed charges of line operators, or equivalently discrete $\theta$ angles of the theory \cite{Gaiotto:2010be,Aharony:2013hda}.  

The aim of this note is to study how these additional data are expressed when the 4d theory considered is a \emph{theory of class $S$}, namely, a theory obtained by compactifying the 6d $\cN{=}(2,0)$ theory on a Riemann surface $C$, and how the S-duality group acts on these data. 
For simplicity, we will only consider Riemann surfaces without punctures. 

The necessity of these additional discrete data arises from the fact that the 6d $\cN{=}(2,0)$ theory does not have a unique partition function on a closed six-dimensional manifold. Rather, it has a partition vector \cite{Aharony:1998qu,Witten:1998wy,Moore:2004jv,Freed:2006yc,Witten:2009at,Henningson:2010rc,Freed:2012bs}, valued in a finite-dimensional vector space with multiple natural bases. When we just compactify this theory on $C$, we get a 4d theory with a partition vector; its components with respect to a chosen basis correspond to partition functions with fixed discrete fluxes through various cycles of the 4d spacetime. 
To obtain a 4d theory with a partition function with Hamiltonian interpretation, we need to sum over the flux sectors in a consistent way. There are multiple ways to achieve this, and this is where the additional discrete data come in.

Let us consider the 6d theory of type $\fg$, and 
 denote by 
 $\cC$ the center of the simply-connected group $G_\text{simp}$ of type $\fg$. 
When the Riemann surface $C$ is $T^2$, the corresponding class $S$ theory is 
the $\cN{=}4$ super Yang-Mills theory with gauge algebra $\fg$.
In addition to the choice of the precise gauge group $G$ whose Lie algebra is $\fg$,
we need to specify a discrete theta angle when $G$ is not simply connected. 
As discussed in \cite{Aharony:2013hda}, these data are equivalent to the specification of maximal set of mutually local discrete electric and magnetic charges of line operators, and is given by a maximal sublattice of $\cC\times \cC$ compatible with the Dirac quantization conditions. 
We identify it with a maximal isotropic sublattice $L$ of $H^1(C,\cC)\simeq \cC\times \cC$, where the isotropy means that the charges are mutually local.
We will also find a formula for the 4d partition function in terms of $L$.
This will reproduce the Vafa-Witten formula \cite{Vafa:1994tf} of the action of the S-duality on the partition function, in a way manifestly generalizable to the whole theories of class $S$.  

By using a Riemann surface $C$ of genus $g$ instead of the torus, we obtain a class $S$ theory. We assume $C$ does not have punctures. Then the corresponding class $S$ theory has $\fg^{3g-3}$ as the gauge algebra, and has as matter contents $2g-2$ copies of the so-called $T_\fg$ theory.  Here, the $T_\fg$ theory is the class $S$ theory for a three-punctured sphere, usually called the $T_n$ theory when $\fg=\mathsf{A}_{n-1}$.
The 6d analysis dictates that the additional discrete data of the class $S$ theory associated to the Riemann surface $C$ are characterized by a maximal isotropic sublattice of $H^1(C,\cC)\simeq \cC^{2g}$. 
We will see that this stems from the fact that there is just one $\cC$ global symmetry  in the $T_\fg$ theory itself, and  that an arbitrary class $S$ theory inherits this global symmetry $\cC$.\footnote{This last point was already found in \cite{Aharony:1998qu} where they say ``this assertion cannot at present be tested independently in any obvious way. It seems like an interesting probe of the inner nature of the still rather mysterious $(0, 2)$ theory.'' It is reassuring that after all these years we understood the nature of the 6d theory at least slightly better.
}  

As a small application, we study how this global $\cC$ symmetry affects the relation of the superconformal index of class $S$ theories and the $q$-deformed Yang-Mills on $C$. We will see that by utilizing this global $\cC$ symmetry of the class $S$ theory, we can obtain $q$-deformed Yang-Mills on $C$ with arbitrary gauge groups belonging to the same $\fg$ with arbitrary discrete torsion. 

\bigskip

The note is organized as follows. In Sec.~\ref{6dreview}, we recall the general structure of the partition function of the 6d $\cN{=}(2,0)$ theory.
In Sec.~\ref{statements}, we summarize how the discrete data of a class $S$ theory are encoded in the 6d language, and describe how the 4d partition function and the 4d Hilbert space are given in terms of the partition function and the Hilbert space associated to the 6d theory. 
In Sec.~\ref{N4}, we study the case of $\cN{=}4$ super Yang-Mills in detail, reproducing the results in \cite{Aharony:2013hda}. 
In Sec.~\ref{classS}, we generalize the discussions to class $S$ theories, by utilizing the global $\cC$ symmetry of the $T_\fg$ theory.  We will find that an arbitrary class $S$ theory always have a global symmetry $\cC$.
In Sec.~\ref{qYM}, we study how this global $\cC$ symmetry affects the equality of the superconformal index of class $S$ theories and the $q$-deformed Yang-Mills on $C$.
We close the note with a brief discussions in Sec.~\ref{conclusions}.
We assume that every four-dimensional manifold we deal with is Euclidean, spin, and does not have torsion in its cohomology,  for simplicity. 

\section{Basics of six-dimensional $\cN{=}(2,0)$ theory}\label{6dreview}

Let us first recall basic properties of 6d $\cN{=}(2,0)$ theories on closed 6d manifolds \cite{Aharony:1998qu,Witten:1998wy,Moore:2004jv,Freed:2006yc,Witten:2009at,Henningson:2010rc,Freed:2012bs}.
\subsection{Discrete three-form fluxes}
\begin{table}
\[
\begin{array}{c||c|c|c|c}
\fg & \mathsf{A}_n & \mathsf{D}_{2n} & \mathsf{D}_{2n+1} & \mathsf{E}_n \\
\cC& \bZ_{n+1} & \bZ_2\times \bZ_2 & \bZ_4 & \bZ_{9-n}
\end{array}
\]
\caption{Lie algebra $\fg$ and the center of simply-connected group of type $\fg$. Note that the formula for $\mathsf{E}_n$ holds for $n=3,\ldots,8$.\label{centers}}
\end{table}

Pick a 6d $\cN{=}(2,0)$ theory of type $\fg$, where $\fg=\mathsf{A}_n$, $\mathsf{D}_n$ or $\mathsf{E}_{6,7,8}$, or a direct sum thereof. 
Given $\fg$, let $\cC$ be the center of the simply-connected group of type $\fg$, as tabulated in Table~\ref{centers}.
There is a natural pairing \begin{equation}
\cC\times \cC \to \U(1).\label{basicpairing}
\end{equation}
For example, when $\cC=\bZ_n$,
the pairing of $x$, $y$ mod $n$ is  given by $e^{2\pi i xy/n}$.
 
We put the 6d theory on a closed six-manifold $X$. The three-form of the 6d theory has discrete fluxes valued in $\cC$ through the three cycles of $X$.
We might be tempted to consider, then, the partition function $Z_a(X)$ of the 6d theory, obtained by fixing the discrete flux to be a given element $a\in H^3(X,\cC)$. 
It is known however that this is not possible, due to the self-duality of the three-form. 
What can be done is as follows. 

The pairing \eqref{basicpairing} on $\cC$ induces an antisymmetric pairing \begin{equation}
H^3(X,\cC)\times H^3(X,\cC) \to \U(1)\label{pairing}
\end{equation} which is non-degenerate. We denote it as \begin{equation}
e^{ i\vev{a,b}}, \qquad \vev{a,b}\in \bR/2\pi\bZ
\end{equation} for $a,b\in H^3(X,\cC)$. For example, when $\cC=\bZ_N$, we have \begin{equation}
\vev{a,b}=\frac{2\pi}{N}\int_X ab.\label{pairingdef}
\end{equation} 

We can not specify the value of the fluxes which have nontrivial pairing under \eqref{pairing}.
Instead, we split \begin{equation}
H^3(X,\cC) = A\oplus B\label{splitting}
\end{equation} where $A$ and $B$ are both isotropic, in the sense that $\vev{a_1,a_2}=\vev{b_1,b_2}=0$ for all $a_{1,2}\in A$ and $b_{1,2}\in B$.  
Then we can fix the discrete flux $a\in A$ and consider the partition function $Z_a$ labeled by $a$.  
Here it is very important that these partition functions are defined relative to the splitting \eqref{splitting}. For example, given another splitting $H^3(X,\cC) = A'\oplus B'$, 
$Z_0$ computed relative this splitting is different from $Z_0$ computed relative to the splitting \eqref{splitting}.
To better understand the situation, it is useful to introduce a vector space $\cZ$ in which we have a partition vector of the 6d theory. A splitting such as \eqref{splitting} equips $\cZ$ with an explicit basis, and $Z_a$ are then the components of the partition vector with respect to this basis. 

\subsection{The partition vector}
Let us implement this procedure concretely.
For $a\in H^3(X,\cC)$, we  define operators $\Phi(a)$ such that \begin{equation}
\Phi(b)\Phi(a)= e^{i\vev{a,b}} \Phi(a)\Phi(b).
\end{equation}
This commutation relation characterizes the uncertainty relation of the discrete $\cC$ fluxes of the self-dual 3-form theory inherent in the 6d $\cN{=}(2,0)$ theory, and is a finite analogue of the standard Heisenberg commutation relation.\footnote{More mathematically, we use the pairing \eqref{pairing} to define the finite Heisenberg group via the extension $$
1\to \U(1)\to \underline{H}^3(X,\cC) \to H^3(X,\cC)\to 0.
$$
Note  that we use additive notation for $a,b\in H^3(X,\cC)$ but multiplicative notation for $\Phi(a),\Phi(b)\in \underline{H}^3(X,\cC)$. Strictly speaking,
 there is no section $\Phi : H^3(X,\cC)\to \underline{H}^3(X,\cC)$ defined on the whole $H^3(X,\cC)$, as is always the case in quantum mechanics due to the operator ordering. We use this slightly wrong but common notation in physics literature.}
It is known that the operators $\Phi(a)$ has a unique  finite-dimensional irreducible representation which we denote by $\cZ$,  see e.g.~\cite{MumfordIII}. The partition vector of the 6d theory takes values in this vector space $\cZ$.

Note that these operators $\Phi(a)$ are not the operators in a 6d quantum field theory in the ordinary sense. They are better thought of as operators in the 7d topological theory whose Hilbert space on a six-dimensional `spatial slice'  gives the space $\cZ$ in which the partition vector takes values.  This is parallel to the situation of the 2d chiral CFTs and the Verlinde line operators: the Verlinde operators are operators acting on the space of conformal blocks, and the space of conformal blocks is the Hilbert space of the 3d topological theory (such as the Chern-Simons theory) on a 2d spatial slice.

One way to construct $\cZ$ and its natural basis is as follows: we split $H^3(X,\cC)$ as in \eqref{splitting}.
Then $\Phi(a)$ for all $a\in H^3(X,\cC)$ commute among themselves, and therefore we can find an explicit basis of $\cZ$ where $\Phi(a)$ are simultaneously diagonalized.
The basis vectors are  given by \begin{equation}
Z_b, \quad b\in B
\end{equation} where we have the actions \begin{equation}
\Phi(a)Z_b = e^{i\vev{a,b}} Z_b,\qquad
\Phi(b)Z_{b'}=Z_{b+b'}.\label{heis}
\end{equation} 
When we vary the metric on $X$, the splitting \eqref{splitting} can be kept constant at least locally. Therefore it makes sense to talk about the dependence of $Z_b$ on the metric. We denote them by $Z_b(X)$, etc.

We can exchange the role of the sublattices $A$ and $B$ in \eqref{splitting}. Then we have $Z^a(X)$ for $a\in A$ where $\Phi(b)$ are diagonalized instead. The bases $\{Z_b\}_{b\in B}$ and $\{Z^a\}_{a\in A}$ are related by a discrete Fourier transformation: \begin{equation}
Z^a = \sum_{b\in B} e^{i\vev{a,b}} Z_b
\end{equation} where we dropped a factor given by a power of $|\cC|$, which are not very important in our analysis in the note. 


\subsection{Hamiltonian interpretation}\label{ham}
If the 6d $\cN{=}(2,0)$ theory were an ordinary quantum field theory, it would associate a Hilbert space to a given a five-dimensional `constant-time-slice' $\tilde X$. 
In our case, the situation is slightly more complicated \cite{Freed:2012bs}.
For simplicity, we only consider the case $X=S^1\times \tilde X$.
we have the canonical splitting \begin{equation}
H^3(S^1\times \tilde X,\cC) = H^3(\tilde X,\cC) \oplus H^2(\tilde X,\cC).\label{ca}
\end{equation} The basis of $\cZ$ where the elements of $H^2(\tilde X,\cC)$ are diagonalized is given by
$Z_v$ for $v\in H^3(\tilde X,\cC)$.
Similarly, the basis of $\cZ$ where the elements of $H^3(\tilde X,\cC)$ are diagonalized is given by
$Z^w$ for $w\in H^2(\tilde X,\cC)$. 
They are related by \begin{equation}
Z^w=\sum_w e^{i\vev{w,v}} Z_v.\label{wv}
\end{equation}
The standard conjecture concerning the 6d $\cN=(2,0)$ theory says that $Z^w$ is essentially given by the path integral of the five-dimensional maximally supersymmetric Yang-Mills theory with gauge group $G_\text{adj}$ of the adjoint type on $\tilde X$,
where the label $w\in H^2(\tilde X,\cC)$ specifies  the Stiefel-Whitney class $w=w_2\in H^2(\tilde X,\cC)$ of the principal $G_\text{adj}$-bundle associated to $\pi_1(G_\text{adj})\simeq \cC$.

As we will see below, to have a consistent Hamiltonian interpretation of class $S$ theory, we require that $Z_v$ with $v\in H^3(\tilde X,\cC)$, instead of $Z^w$ with $w\in H^2(\tilde X,\cC)$, to have the Hamiltonian interpretation: \begin{equation}
Z_v = \Tr_{\cH_v} (-1)^F e^{-\beta H} \label{hamil}
\end{equation} 
where $\beta$ is the circumference of $S^1$, $F$ is the fermion number, $H$ is the Hamiltonian of the system, acting on the Hilbert space $\cH_v$.

\section{Statement of the results}\label{statements}
Let us consider the class $S$ theories obtained by compactifying this theory on a Riemann surface $C$ of genus $g$.
The class $S$ theories consists of $3g-3$ copies of $\fg$ $\cN{=}2$ vector multiplets and $2g-2$ copies of the $T_{\fg}$ theory.
We now need to specify additional discrete data to fully specify the 4d theory.
Instead of giving various arguments and then extracting the results, we present our conclusions in a concise form in this section. The consistency of the statements below will be checked in the sections that follow.
\begin{enumerate}
\item The additional data are the \emph{maximal isotropic sublattice} $L$ of $H^1(C,\cC)$.
Here the isotropy means that for any $l_{1,2}\in L$ we have $\vev{l_1,l_2}=0$.
The sublattice $L$ specifies the set of the allowed discrete charges of the line operators of the theory.
Let us denote such a fully specified theory by $S_\fg(C,L)$. 
\item Given a 4d manifold $Y$, the partition function of this class $S$ theory is given by a unique element $Z_L(Y\times C)$ of the space $\cZ(Y\times C)$, specified by the condition that $Z_L$ is  invariant under the quantized discrete fluxes in $H^2(Y,\cC)\otimes L$: \begin{equation}
\Phi(v)  Z_L = Z_L,  \quad \text{for all $v\in H^2(Y,\cC)\otimes L$.}\label{invariance}
\end{equation}
\item A class $S$ theory (for a Riemann surface $C$ of genus $g\ge 2$ without any puncture) always has the global symmetry $\cC$. Therefore, on a four-dimensional manifold with nontrivial $\pi_1$, we can introduce a background flat connection for this global symmetry $\cC$. Equivalently, we can gauge a subgroup of $\cC$.
\item Given a 3d `constant-time-slice' $\tilde Y$, the Hilbert space $\cH_{S_\fg(C,L)}(\tilde Y)$ of the class $S$ theory is given by\footnote{This statement needs a slight modification when $H_\bullet(\tilde Y,\bZ)$  has torsion. In that case, there are subtle shifts in $v$ to sum over, presumably due to the discrete charge flux generated by the geometry, see \cite{Razamat:2013opa}.} \begin{equation}
\cH_{S_\fg(C,L)}(\tilde Y)= \bigoplus_{
\begin{subarray}{l}
v \in H^2(\tilde Y,\cC)\otimes L,\\
k\in H^3(\tilde Y,\cC)\simeq \cC
\end{subarray}
} \cH_{v,k}(\tilde Y\times C).
\label{Hilbert}
\end{equation} Here, for concreteness, we considered $\tilde Y$ with trivial $\pi_1$, and a class $S$ theory whose global $\cC$  symmetry is not gauged.  The spaces $\cH_{v,k}$ on the right hand side are the Hilbert spaces given by the 6d theory, and $v$ measures the electric and magnetic fluxes through the two cycles of $\tilde Y$, and $k$ specifies the charge under the global $\cC$ symmetry of the class $S$ theory. Note that the 6d theory itself defines $\cH_{v,k}$ for arbitrary $v\in H^2(\tilde Y,\cC)\otimes H^1(C,\cC)$, and we take the direct sum over a specific subset determined by $L$.
\end{enumerate}
These properties are stated without any reference to any decomposition of $C$ into pants. Therefore the action of the S-duality, i.e.~the mapping class group of $C$, is completely transparent in this formulation. But the relation to the choice of the gauge group and of the discrete theta angles is made somewhat obscure. 
We will study these points in the rest of the note. 

Before proceeding, we stress that this additional data $L\subset H^1(C,\cC)$ are specified \emph{in addition to} a decomposition into pants when we talk about a weakly-coupled duality frame, although a decomposition into pants determines a natural isotropic decomposition $H^1(C,\bZ)=A\oplus B$ and therefore a natural maximally isotropic sublattice $A\otimes \cC$ and $B\otimes \cC$.  Surely $L=A\otimes \cC$ and $L=B\otimes \cC$ are two natural choices of the additional data given a weakly-coupled frame; but they are not all.

\section{$\cN{=}4$ super Yang-Mills}\label{N4}
\subsection{$\SU(N)$ and $\SU(N)/\bZ_N$}
Now, let us consider the case where $C=T^2$. Then the low-energy limit of the four-dimensional system is described by the $\cN{=}4$ super Yang-Mills theory with gauge algebra $\fg$. 
The group $H^2(C,\cC)=\cC\times \cC$ can be identified with the discrete electric and magnetic charges of the Wilson, 't Hooft or in general dyonic line operators of $\cN{=}4$ super Yang-Mills theory.
The pairing \begin{equation}
H^2(C,\cC) \times H^1(C,\cC) \to \U(1) 
\end{equation} is the modulo one of the Dirac quantization pairing. 
An isotropic sublattice $L$ of $H^1(C,\cC)$ 
is such that for any $l_{1,2}\in L$ we have $\vev{l_1,l_2}=0$, i.e.~they satisfy the Dirac quantization law.
A maximal isotropic sublattice $L$ is an isotropic sublattice to which we can add any more element preserving the isotropy.
Therefore, such an $L$ can naturally be identified with the allowed set of charges of line operators in a consistent theory. 

For simplicity we consider $\fg=\mathsf{A}_{N-1}$ and therefore $\cC=\bZ_N$. Generalization to other cases is immediate. Let us define a basis of $H^1(C,\cC)$ by saying that every element can be written as fix a splitting \begin{equation}
eW+mH\in H^1(C,\cC)
\end{equation} where $e,m=1,\ldots, N$. This picks a particular duality frame, such that $W$ is the Wilson line in the fundamental representation and $H$ is the 't Hooft line with the minimal magnetic charge. 
Consider a 4d manifold $Y$. For simplicity let us assume $H^1(Y)$ and $H^3(Y)$ are trivial.
Then, we can  split $H^3(Y\times C,\cC)$ as follows: \begin{equation}
H^3(Y\times C,\cC) = (H^2(Y,\cC) \otimes W) \oplus 
(H^2(Y,\cC) \otimes H).
\end{equation} The first term and the second term on the right hand side 
correspond to electric and magnetic fluxes through two-cycles of $Y$, respectively.
We use a basis of $\cZ(Y\times C)$ where the part $H^2(Y,\cC) \otimes W$ is diagonalized.
Equivalently, we consider a partition vector whose components are given by \begin{equation}
Z_v:=Z_{v\otimes H} (Y\times C), \quad v\in H^2(Y,\cC).
\end{equation}
We identify $Z_v$ to be the partition function of $\cN{=}4$ super Yang-Mills based on a principal bundle of $G_\text{adj}=\SU(N)/\bZ_N$, with the condition that the Stiefel-Whitney class of the bundle associated to $\pi_1(G_\text{adj})\simeq \cC$ is fixed to $v$.

When $L=\{eW\}$, the element in $\cZ(Y\times C)$ invariant under the quantized action of $H^2(Y,\cC)\times L$ is clearly just $Z_0$.  This is, up to a constant multiple, the partition function of $\cN{=}4$ super Yang-Mills with gauge group $\SU(N)$, because trivial Stiefel-Whitney class $0\in H^2(Y,\cC)$ means that the gauge bundle can be lifted to an $\SU(N)$ bundle. 
We note that $\{eW\}$ is the maximal set of the charges of the allowed line operators of the theory with gauge group $\SU(N)$. 

When $L=\{mH\}$, the element in $\cZ(Y\times C)$ invariant under the quantized action of $H^2(Y,\cC)\times L$ is clearly just \begin{equation}
\sum_v Z_v,
\end{equation} recall the explicit action of the Heisenberg group in \eqref{heis}.
 This is the partition function of $\cN{=}4$ super Yang-Mills with gauge group $\SU(N)/\bZ_N$ with zero theta angle. Indeed $\{mH\}$ is the maximal set of the charges of the allowed line operators of the theory with gauge group $\SU(N)/\bZ_N$, and we summed over all possible Stiefel-Whitney classes $v$. 

The choices $L=\{eW\}$ and $L=\{mH\}$ can be exchanged by changing the complex structure of the torus as $\tau \mapsto -1/\tau$, and therefore we have \begin{equation}
Z_0(\tau)=\sum_v Z_v(-\frac{1}\tau).
\end{equation}
This relation of the partition functions $Z_0$ and $\sum_v Z_v$ goes back to \cite{Vafa:1994tf}. The explanation using discrete Fourier transform was already essentially given in \cite{Witten:2009at}. Here,  we identified the partition function via the condition of invariance under $H^2(Y,\cC)\otimes L$. This is also essentially done e.g.~in \cite{Aharony:1998qu}.

\subsection{General case}
Let us now consider a general maximal sublattice $L\subset \cC\times \cC$,
specifying the set of charges of mutually local line operators.
We would like to determine the element $Z_L$ of $\cZ(Y\times C)$ invariant under the quantized action of $H^2(Y,\cC)\otimes L$, as a linear combination of $Z_v$ where $v\times H\in H^2(Y,\cC)\otimes H^1(C,\cC)$ specifies the Stiefel-Whitney class of the gauge bundle. 
This gives an explicit formula of the partition function of the $\cN=4$ super Yang-Mills with the set of line operators being $L$, in terms of a summation over the topological class of gauge bundles with an explicit phase factor.

For concreteness, we again restrict to the case $\fg=\mathsf{A}_{N-1}$ and therefore $\cC=\bZ_N$.  Let $L\cap \{eW\}_{e=1,2,\ldots}= \{kiW\}_{i=1,2\ldots}$, where $k$ is an integer dividing $N$. 
Then an element in $L$ of the form $mH +nW$ with minimal positive $m$ has $m=N/k$ from maximality of $L$. We have the choice $n=0,1,2,\ldots,k-1$. 
Then  $L$ is generated by $kW$ and $mH+nW$ with $m=N/k$.

We now need to construct $\Phi(a)$ for all $a\in H^2(Y,\cC)\otimes L$, such that
$\Phi(a+b)=\Phi(a)\Phi(b)$. 
To do this, we need to specify
$\Phi(w\otimes kW)$ and $\Phi(w\otimes (mH+nW))$.  The representation of the former is given already in \eqref{heis}, \begin{equation}
\Phi(w\otimes kW) Z_{v}=e^{ik\vev{w,v}} Z_{v}.\label{999}
\end{equation} 

The representation of the latter needs some more work.
In $\Phi(w\otimes (mH+nW))$, we can consider $w$ as an element of $H^2(Y,\bZ_k)$.
We distinguish the pairing of $H^2(Y,\bZ_k)$ by writing it as $\vvevv{\cdot,\cdot}$, defined as in \eqref{pairingdef}, 
and keep using $\vev{\cdot,\cdot}$ for the pairing of $H^2(Y,\bZ_N)$. 
Note that for $w\in H^2(Y,\bZ_k)$, we have $mw\in H^2(Y,\bZ_N)$,
and our normalizations are that $\vev{mw,mw}=m\vvevv{w,w}$.
We find that  the following satisfies all the requirements: \begin{equation}
\Phi(w \otimes (mH+nW)) Z_v = e^{inm\vev{w,v} +in\vvevv{w,w}/2} Z_{v+mw}.\label{000}
\end{equation}
Here,  the phase  \begin{equation}
\frac{\vvevv{w,w}}{2}= \frac{2\pi }{k}\int_Y \frac{w\cdot w}{2}\label{___}
\end{equation} 
is more properly understood to be given using the Pontrjagin square $\mathfrak{P}(w)$, if necessary; we will continue to use our informal notation. In our convention,   the quantity $\int_Y w\cdot w/2$ on  a spin manifold is well defined as an integer modulo $k$. For more on the Pontrjagin square, see \cite{Henningson:2010rc,Kapustin:2013qsa}.

The invariance under \eqref{999} means that  only those $Z_v$ with $kv=0$ modulo $N$ appear in $Z_L$. 
Equivalently, we have $v\in H^2(Y,\bZ_k)\subset H^2(Y,\bZ_N)$.
The invariance under \eqref{000} then requires that we have \begin{equation}
Z_L= \sum_{v\in H^2(Y,\bZ_k)} e^{i n\vvevv{v,v}/2} Z_v.\label{generalN4part}
\end{equation} 

The restriction $v\in H^2(Y,\bZ_k)\subset H^2(Y,\bZ_N)$ in the sum means that  this is a partition function of $\cN{=}4$ gauge theory with gauge group $\SU(N)/\bZ_k$, and the phase $in\vvevv{v,v}/2$ 
is  exactly the structure found in \cite{Aharony:2013hda} for the $\cN{=}4$ gauge theory with the set $L$ of line operators. 

\section{Class $S$ theories}\label{classS}
The discussion in the last section did not depend much on the particular choice of $C=T^2$, and therefore, it can naturally be generalized to other $C$ of general genus  $g\ge 1$. 
For simplicity we assume there is no puncture and no outer-automorphism twist lines on $C$, and 
let us consider a theory of class $S$ obtained by compactifying the 6d $\cN{=}(2,0)$ theory of type $\fg$ on $C$. 

The only point which needs a further discussion is the identification of the $H^2(C,\cC)\simeq \cC^{2g}$ as the possible label of the discrete charges of line operators.   
Before proceeding, we note that
for the class $S$ theories of type $\mathsf{A}_1$,
 the analysis of the charges of all possible line operators was carried out in a beautiful paper \cite{Drukker:2009tz}, and then the possible choices of maximally mutually-local subset of charges were discussed in \cite{Gaiotto:2010be}. What is discussed below is a natural generalization of their discussions.

Let us fix a pants decomposition of $C$. Correspondingly, the theory has a realization as an $\cN{=}2$ gauge theory with gauge algebra $\fg^{3g-3}$ coupled to $2g-2$ copies of the $T_\fg$ theory, i.e.~the class $S$ theory corresponding to a three-punctured sphere. To analyze the combined system, we need to recall some properties of the $T_\fg$ theory.  

\subsection{Action of the center on the $T_\fg$ theory}
The theory $T_\fg$  has $\fg^3$ flavor symmetry. The operator content of $T_\fg$ is not fully known, although its superconformal index has a rather-well-established conjectural form, given in \cite{Gadde:2011ik,Gadde:2011uv}.  There, it was found that every operator which contributes to the superconformal index is in a representation $R_1\otimes R_2\otimes R_3$ of $\fg\oplus \fg \oplus \fg$, \emph{such that the action of\/ $\cC$ on $R_{1,2,3}$ are the same.}
Equivalently, the $T_\fg$ theory has a flavor symmetry $G_\text{simp}^3$, where $G_\text{simp}$ is the simply-connected Lie group with the Lie algebra $\fg$, but the center $\cC^3$ does not all act independently on the theory. Rather, there is a natural map 
\begin{equation}
\cC\times \cC\times \cC \to \cC_\text{tri-diag} \label{tridiag}
\end{equation} given by $(a,b,c)\mapsto abc$,
and only $\cC_\text{tri-diag}$ acts faithfully on the $T_\fg$ theory.

It seems natural to assume that these statements on the center charges of the operators of $T_\fg$ theory hold including non-BPS operators. As we will see, this assumption leads to a consistent interpretation of the properties of the class $S$ theory, and it seems difficult (at least to the author) to add non-BPS operators which do not satisfy this property, still preserving the overall consistency. 
When $\fg=\mathsf{A}_1$, the $T_{\mathsf{A}_1}$ theory is a theory of eight free chiral multiplets $Q_{iau}$, where $i,a,u=1,2$ are the $\SU(2)$ flavor symmetry indices for $\SU(2)^3$. In this case, it is easy to see that the center $\bZ_2$ of any of the three $\SU(2)$ just multiplies $Q_{iau}$ by $\pm 1$, confirming the assumption above.
When $\fg=\mathsf{A}_2$, the $T_{\mathsf{A}_2}$ theory is believed to have an enhanced symmetry $\mathsf{E}_6$. The assumption above translates in this language that all operators are representations of the adjoint form $\mathsf{E}_6/\bZ_3$.

Before proceeding, we note that the statement that only $\cC_\text{tri-diag}$ acts faithfully only applies to point operators of the $T_\fg$ theory.  It is easy to consider external line operators on which the full action of $\cC\times \cC\times \cC$ can be distinguished: we just have to consider a pure flavor Wilson line operator associated to one of the $G$ symmetries. For example, in $T_{\mathsf{A}_1}$  theory, we can consider a Wilson line in the fundamental representation of each of the three $\SU(2)$ symmetries.
Similarly, we expect that there are external line operators of the $T_{\mathsf{A}_2}$ theory which do not transform under the adjoint form $\mathsf{E}_6/\bZ_3$, but under the simply-connected version $\mathsf{E}_6$ \cite{TachikawaWatanabe}.

\subsection{Discrete charge lattice of class $S$ theories}
Now, let us come back to the study of the class $S$ theory one obtains by compactifying the 6d theory of type $\fg$ on a Riemann surface $C$ of genus $g\ge 2$ without any punctures. 
We have the $\fg^{3g-3}$ gauge multiplets, coupled to $2g-2$ copies of the $T_\fg$ theory. 

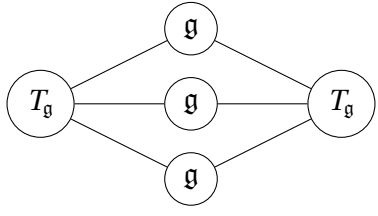
\begin{figure}
\centering
\begin{tikzpicture}
\node[draw,circle] (1) at (-2,0) {$T_\fg$};
\node[draw,circle] (2) at (2,0) {$T_\fg$};
\node[draw,circle] (A) at (0,1) {$\fg$};
\node[draw,circle] (B) at (0,0) {$\fg$};
\node[draw,circle] (C) at (0,-1) {$\fg$};
\draw (1)--(A)--(2);
\draw (1)--(B)--(2);
\draw (1)--(C)--(2);
\end{tikzpicture}
\caption{A weakly-coupled frame of a class $S$ theory for a genus-2 Riemann surface.\label{genus2}}
\end{figure}

If we have the $\fg^{3g-3}$ gauge multiplets in isolation, we have a lattice $E_\text{naive}=\cC^{3g-3}$ of discrete electric charges and 
a lattice $M_\text{naive}=\cC^{3g-3}$ of discrete magnetic charges. 
In the last subsection, we argued that a single $T_\fg$ theory has a single $\cC$ global symmetry.  With $2g-2$ copies, we have the flavor symmetry $\cC^{2g-2}$.
Then, we have a natural map \begin{equation}
\sigma: \cC^{3g-3} \to  \cC^{2g-2}.\label{centeraction}
\end{equation} controlling how the center $\cC^{3g-3}$ acts on the copies of the $T_\fg$ theory.
A crucial point is that the image of this action is $\cC^{2g-3}$.  For example, consider the case $g=2$. Let us take a duality frame where three $\fg$  couple to a diagonal combination of one $\fg$ from one $T_\fg$ and another $\fg$ from another $T_\fg$, see Fig.~\ref{genus2}. Then, the map $\sigma$ is given by \begin{equation}
\sigma:\cC^{3}\ni (a,b,c) \mapsto (abc,(abc)^{-1}) \in \cC^2 
\end{equation} and therefore the subgroup\begin{equation}
(a,1)\in \cC^2
\end{equation} is not gauged. 

Therefore, we have the  sublattice  
\begin{equation}
\Lambda_{g} \simeq \cC^{2g-3} \subset E_\text{naive} \simeq \cC^{3g-3}
\end{equation}
characterizing the center charges of the dynamical operators coming from the copies of the $T_\fg$ theory. 
Due to the screening by the dynamical operators of charges $\Lambda_g$,
the lattice of the discrete electric external line operators is now \begin{equation}
E=E_\text{naive}/\Lambda_g\simeq \cC^g.
\end{equation} In order to satisfy the Dirac quantization condition with respect to the dynamical operators of charges $\Lambda_g$, the lattice of the discrete magnetic external line operators is now \begin{equation}
M=\{m\in M_\text{naive} \mid \vev{m,\Lambda_g}=0\}\simeq \cC^g.
\end{equation}
Therefore, we naturally have the identification \begin{equation}
H^1(C,\cC)\simeq \cC^{2g} \simeq E\oplus M.\label{foo}
\end{equation} Note  that the relation \eqref{tridiag} can be thought of describing the $H^1$ of a sphere minus three punctures, if we identify each $\cC$ with the dual of $S^1$ around one puncture. Then, the splitting of $H^1(C,\cC)$ into $E$ and $M$ in  \eqref{foo} is the geometrically natural one associated to the pants decomposition.

Now, we can repeat the analysis given in Sec.~\ref{N4} for the $\cN=4$ super Yang-Mills when $C=T^2$ almost verbatim. 
The maximal set of mutually local line operators is given by a maximally isotropic sublattice $L\subset H^1(C,\cC)$.\footnote{For the 6d theory of type $\fg=\mathsf{A}_1$, this statement was originally found in \cite{Gaiotto:2010be}. There, the isotropy was stated as the condition that the chosen one-cycles on $C$ should have even intersection numbers.}
The partition function of the 4d class $S$ theory on a four-manifold $Y$ with trivial $\pi_1$ is then given by the essentially unique vector in $\cZ$, invariant under $\Phi(v)$ for all $v \in H^2(Y,\cC)\otimes L$.

Recall the action \eqref{centeraction} of the center of the gauge groups to the global symmetries $\cC^{2g-2}$ of copies of the $T_\fg$ theory. The image is $\cC^{2g-3}$, and therefore we see that the class $S$ theory associated to a Riemann surface of  genus $g\ge 2$ without any puncture has the global symmetry $\cC$, independent of the choice of the set $L$ of allowed charges of line operators.  In a weakly-coupled frame, the action of this universal global $\cC$ symmetry to  act on one of the $T_\fg$ theory nontrivially, and acts on all the other copies of the $T_\fg$ theories trivially.  
The existence of  global symmetry $\cC$ for any class $S$ theory was already pointed out in \cite{Aharony:1998qu}.

It is of course possible to gauge a subgroup of this global $\cC$ symmetry further, and regard the resulting theory as a new class $S$ theory. In the rest of the paper, however, we stick to the convention where this global $\cC$ symmetry is not considered to be gauged; we will still utilize background $\cC$ gauge fields.

A priori, a strongly-coupled theory such as the $T_\fg$ theory can have intrinsic additional discrete data corresponding to the choice of the allowed set of external line operators, just as the $\cN{=}4$ super Yang-Mills theory with a given Lie algebra $\fg$ had. 
The $T_\fg$ theory has one such additional data, which are rather trivial: we can gauge a subgroup $\Gamma$ of the tridiagonal center symmetry $\cC_\text{tridiag}$ acting on $T_\fg$. The partition function of the $T_\fg$ theory with different choices of $\Gamma$ will then be different on a non-simply-connected manifold. But this is a rather trivial additional data, related purely to the flavor symmetry. 

The $T_\fg$ theory does not seem to have any more discrete data in addition to this.
If it really had such additional data, that would give an additional term on the right hand side of \eqref{foo}, thus ruining the overall structure of the 6d $\cN{=}(2,0)$ theory recalled in Sec.~\ref{6dreview} and the discussions in Sec.~\ref{N4}. Therefore, it is strongly likely that \emph{the $T_\fg$ theory does not have a choice of the allowed set of external line operators,} except those coming from the gauging of the center flavor symmetry $\cC$.

\subsection{Hamiltonian interpretation of a class $S$ theory}
Let us now consider the Hamiltonian interpretation when the four-manifold $Y$ has the form $Y=S^1\times \tilde Y$. 
We assume that $\tilde Y$ has trivial $\pi_1$ for simplicity. 
To get the Hamiltonian interpretation, we need to  use the basis of $\cZ$ associated to the splitting \begin{equation}
H^3(Y\times C)=H^3(\tilde Y\times C)\oplus H^2(\tilde Y\times C)\label{u}
\end{equation} and simultaneously diagonalizing $\Phi(x)$ in $x\in H^2(\tilde Y\times C)$ in \eqref{u}.  
Therefore the elements of the partition vector is given by $Z_{v,k}$ where
 \begin{equation}
(v,k)\in [H^2(\tilde Y)\otimes H^1(C)]\oplus H^3(\tilde Y)=H^3(\tilde Y\times C).
\end{equation}

On this basis, we need to construct the quantized action of 
\begin{equation}
H^2(Y,\cC)\otimes L= (H^2(\tilde Y,\cC)\otimes L)\oplus (H^1(\tilde Y,\cC)\otimes L).
\end{equation} Calling an element $v\oplus w$, 
the action \eqref{000} is now given by\footnote{This is true only when $H_\bullet(\tilde Y,\bZ)$ has no torsion. In general there can be a term in the exponent proportional to the torsion part of $w$.  } \begin{equation}
\Phi(v\oplus w) Z_{v',k}=e^{2\pi i \vev{v,w}} Z_{v+v',k}.
\end{equation} 
Then the combination which is invariant under these operators is \begin{equation}
\sum_{
\begin{subarray}{l}
v \in H^2(\tilde Y,\cC)\otimes L,\\
k\in H^3(\tilde Y,\cC)\simeq \cC
\end{subarray}
}  Z_{v,k},\label{z}
\end{equation} and therefore the Hilbert space associated to $\tilde Y$ is given by \begin{equation}
 \bigoplus_{
\begin{subarray}{l}
v \in H^2(\tilde Y,\cC)\otimes L,\\
k\in H^3(\tilde Y,\cC)\simeq \cC
\end{subarray}
} \cH_{v,k}(\tilde Y\times C).\label{hzw}
\end{equation} This is the statement in \eqref{Hilbert}.

The additional summation over $k\in H^3(\tilde Y,\cC)\simeq \cC$ can naturally be thought of as a summation over the charge sectors under the global symmetry $\cC$ which exists for any class $S$ theory. This feature will be elaborated further in the next section.

\subsection{The 3d Coulomb branch}\label{hitchin}
Before proceeding, let us now discuss the 3d Coulomb branch of the moduli space of a class $S$ theory $S_\fg[C,L]$ compactified on a circle $S^1$.  This problem was already analyzed in \cite{Gaiotto:2010be} for the case $\fg=\mathsf{A_1}$ and  there is no essential change in this general case.

Consider the moduli space $\cM$ of the Hitchin system on a Riemann surface $C$ of genus $g$ without puncture. We let the gauge group of the Hitchin system to be $G_\text{simp}$, the simply connected group associated to the Lie algebra $\fg$. 
There is the Hitchin fibration \begin{equation}
p: \cM \to \cM_\text{4d Coulomb}
\end{equation} where $\cM_\text{4d Coulomb}$ is the 4d Coulomb branch. The generic fiber  is $T^{2g\rank \fg}$ and is an Abelian variety. 

There is a natural action of $H^1(C,\cC)$ on $\cM$, which commutes with $p$, constructed as follows. We identify an element  $l\in H^1(C,\cC)$ with a flat $\cC$ bundle $\cL(l)$ over $C$. Then we can ``tensor'' $\cL(l)$ to the $G_\text{simp}$-bundle in the the Hitchin system, by multiplying the transition functions. This can be consistently done, because $\cC$ is the center of $G_\text{simp}$.  This operation clearly commutes with the Hitchin fibration $p$. 
Then $\cM/H^1(C,\cC)$ is the moduli space of the $G_\text{adj}$-Hitchin system on $C$, where  the topological class of the $G_\text{adj}$-bundle is assumed to be trivial.

Given a maximally isotropic sublattice $L\subset H^1(C,\bC)$, we can instead take the quotient \begin{equation}
p: \cM/L \to \cM_\text{4d Coulomb}.
\end{equation} Then we identify $\cM/L$ as the 3d Coulomb branch of the class $S$ theory $S_\fg[C,L]$. 

Presumably, the fiber of  $\cM/L$ is a principally-polarized Abelian variety. More generally, the fibration of the Donagi-Witten integrable system of an $\cN=2$ supersymmetric theory, once the maximal set $L$ of the mutually local line operators is fixed, will be a principally-polarized Abelian variety. This point needs to be studied in more detail.


\section{Superconformal index and $q$-deformed Yang-Mills}\label{qYM}
The global symmetry $\cC$ which exists for any class $S$ theory can be used to refine the relation of the superconformal index of class $S$ theories and 2d $q$-deformed Yang-Mills originally found in \cite{Gadde:2011ik,Gadde:2011uv}. We will see below that by utilizing $\cC$ we can have 2d $q$-deformed Yang-Mills with non-simply-connected gauge group with or without discrete torsion.

The superconfomal index of a 4d $\cN=2$ theory is the partition function on $S^3\times S^1$, with an appropriate choice of background fields to preserve supersymmetry. Here, we only consider the simplest case with one parameter $q$, where we have \begin{equation}
Z(S^3\times S^1)=\Tr_{\cH} (-1)^F q^{E-R}
\end{equation} where $\cH$ is the Hilbert space on the $S^3$, $F$ is the fermion number, $E$ is the energy (or the scaling dimension of the operator under the state-operator correspondence), and $R$ is the $\SU(2)$ R-symmetry normalized to take $\pm1/2$ in the fundamental representation. 

Consider a class $S$ theory $S_\fg[C_g,L]$ obtained by compactifying the 6d $\cN=(2,0)$ theory of type $\fg$ on a Riemann surface $C_g$ of genus $g\ge 2$  without any puncture. 
As $S^3\times S^1$ does not have two-cycles, the partition function does not depend on the choice of the maximal isotropic sublattice $L$, and we have \cite{Gadde:2011ik,Gadde:2011uv}\begin{equation}
Z_{S_\fg[C_g,L]}(S^3\times S^1)
=\frac{1}{K_\fg(q){}^{2g-2}}\sum_{\lambda} \frac{1}{(\dim_q \lambda)^{2g-2}},
\end{equation}  where $K_\fg(q)$ is a certain prefactor,
the summation is over the irreducible representations $\lambda$ of $\fg$,
and $\dim_q \lambda$ is the quantum dimension of $\lambda$. 
Up to a prefactor, this is the partition function of the $q$-deformed Yang-Mills 
on $C_g$ with gauge group $G_\text{simp}$, where $G_\text{simp}$ is the simply-connected one associated to $\fg$.

This formula can be derived from the superconformal index of the $T_\fg$ theory \begin{equation}
Z_{T_\fg}(S^3\times S^1)(a,b,c)=
\Tr_{\cH}  (-1)^F q^{E-R} abc
=\frac{K(a)K(b)K(c)}{K_\fg(q){}^{2g-2}}\sum_{\lambda} \frac{\chi_\lambda(a)\chi_\lambda(b)\chi_\lambda(c)}{(\dim_q \lambda)^{2g-2}},
\end{equation} where $(a,b,c)\in G_\text{simp}^3$ are the exponentiated chemical potentials for the flavor symmetry. 
The numerator has factors $\chi_\lambda(a)$ which is the character of $a$ in the representation $\lambda$, and also prefactors $K(a)$ which purely consists of characters of tensor powers of adjoint representations. 

Note that the tri-diagonal center symmetry $\cC_\text{tridiag}$, \eqref{tridiag}, is manifest.
Let us define the pairing $(\gamma,\lambda)$ of an element $\gamma \in \cC\subset G_\text{simp}$ 
and an irreducible representation $\lambda$ of $G_\text{simp}$ by saying that $\gamma$
acts by the multiplication by a phase $e^{2\pi i(\gamma,\lambda)}$ on the irreducible representation $\lambda$.
Then 
\begin{equation}
Z_{T_\fg}(S^3\times S^1)(\gamma)=
\Tr_{\cH}  (-1)^F q^{E-R} \gamma
=\frac{K(a)K(b)K(c)}{K_\fg(q){}^{2g-2}}\sum_{\lambda} e^{2\pi i(\gamma,\lambda)} \frac{\chi_\lambda(a)\chi_\lambda(b)\chi_\lambda(c)}{(\dim_q \lambda)^{2g-2}},
\end{equation}
where we now regard $\gamma $ to be an element in 
\begin{equation}
\gamma\in \cC_\text{tridiag} \subset \cC^3\subset G_\text{simp}^3.
\end{equation}

From this, we find that the superconformal index of a class $S$ theory $S_\fg[C_g,L]$ for a genus $g$ surface $C$ without any puncture, with a twist $\gamma$ around $S^1$ in the universal global symmetry $\cC$, is given by \begin{equation}
Z_{S_\fg[C_g,L]}(S^3\times S^1)(\gamma)
=\frac{1}{K_\fg(q){}^{2g-2}}\sum_{\lambda} e^{2\pi i(\gamma,\lambda)}\frac{1}{(\dim_q \lambda)^{2g-2}}.
\end{equation}

Then, we find \begin{equation}
\sum_\gamma Z_{S_\fg[C_g,L]}(S^3\times S^1)(\gamma)
=\frac{1}{K_\fg(q){}^{2g-2}}\sum_{(\gamma_0,\lambda)=0\,\text{mod}\, 1} \frac{1}{(\dim_q \lambda)^{2g-2}}.
\end{equation} Now, the label $\lambda$ runs over the irreducible representation of $G_\text{adj}$, i.e.~the adjoint form associated to the Lie algebra $\fg$. 
This gives the partition function of the $q$-deformed Yang-Mills theory with gauge group $G_\text{adj}$.

We can also include additional phase factor. For concreteness, let us take $\fg=\mathsf{A}_{N-1}$. Denote by $\gamma_0$ the generator of $\cC=\bZ_N$.
We then have\begin{equation}
\sum_n e^{2\pi i kn/N} Z_{S_\fg[C_g,L]}(S^3\times S^1)(\gamma_0{}^n)
=\frac{1}{K_\fg(q){}^{2g-2}}\sum_{(\gamma_0,\lambda)= k/N\, \text{mod}\, 1} \frac{1}{(\dim_q \lambda)^{2g-2}}.
\end{equation} Now,  the summation is over the irreducible representation of $G=\SU(N)$ whose $N$-ality is $k$ mod $N$.
This gives the partition function of the $q$-deformed Yang-Mills theory with gauge group $G_\text{adj}=\SU(N)/\bZ_N$, with an additional phase factor \begin{equation}
\frac{2\pi i k}N \int_C w_2\label{dt}
\end{equation} in the Lagrangian, where $w_2\in H^2(C,\cC)$ is the Stiefel-Whitney class of the $G_\text{adj}$ bundle over the Riemann surface $C$. 

Note that by including the phase factor $e^{2\pi i k n /N}$ on the left hand side, we are projecting down to a subspace of the Hilbert space with a fixed charge $k$ under the global symmetry $\cC=\bZ_N$. This corresponds to restricting to a single summand of $k\in H^3(\tilde Y,\cC)$ in the general expression \eqref{hzw} for the Hilbert space and the partition function \eqref{z}, where $\tilde Y=S^3$ in this setup. 

The appearance of $G_\text{adj}$  as the 2d gauge group can be understood as in \cite{Fukuda:2012jr}, where the $S^1$ direction is compactified first. In this approach, we have the 5d maximally supersymmetric Yang-Mills on $\tilde X=\tilde Y\times C$. 
As explained in Sec.~\ref{ham} and in particular \eqref{wv},  computing the partition function fixing an element $k\in H^3(\tilde Y,\cC)$  means that weighing the partitinon function of the 5d super Yang-Mills with gauge group $G_\text{adj}$ with nontrivial $w_2\in H^2(C,\cC)$, exactly with the weighting factor \eqref{dt}. Reducing the 5d Yang-Mills on $S^3$, we obtain a 2d Yang-Mills with gauge group $G_\text{adj}$. 

So far we showed how to obtain $G_\text{adj}$ as the 2d gauge group. It is clear that we can extend the construction in this section to have the $q$-deformed Yang-Mills on $\cC$ for the arbitrary gauge group $G$ whose Lie algebra is $\fg$ together with arbitrary discrete torsion \eqref{dt}.

\section{Conclusions and future directions}\label{conclusions}
To fully specify a gauge theory, we need to specify the set of allowed charges of line operators, or equivalently the global structure of the gauge group together with the discrete theta angles, as first found in \cite{Gaiotto:2010be,Aharony:2013hda}. 
In this short note, we studied how these data are encoded in the 6d $\cN{=}(2,0)$ theory, when the gauge theory we consider is a theory of class $S$ in 4d. The results are summarized in Sec.~\ref{statements} and we do not repeat them here. 

Many points need to be clarified further. We list some of them below: \begin{itemize}
\item To understand more fully the behavior of the partition vector of the 6d $\cN=(2,0)$ theory of type $\fg$.  It is known that the 6d theory of type $\U(N)$ has an honest partition function, and the rather subtle behavior of the partition vector of the 6d theory of type $\SU(N)$ arises from trying to decouple the Abelian $\U(1)$ part, see e.g.~\cite{Moore:2004jv}. 
Also, when the center $\cC$ of the simply-connected group $G$ associated to the Lie algebra $\fg$ is of the form $\cC\simeq \bZ_{n^2}$, we can choose a natural maximally isotropic sublattice of $H^3(X,\cC)$ for arbitrary closed six-manifold $X$ by requiring the flux to be annihilated by multiplication by $n$. Then we have a genuine quantum field theory in 6d.\footnote{The author thanks O. Aharony and N. Seiberg for explaining this fact to him. } 
Starting from a genuine 6d quantum field theory using these constructions might shed a new light on the behavior of the 4d theories.
\item To extend the analysis in this note to the class $S$ theory associated to the Riemann surfaces with punctures. To do this, we need to understand the behavior of the partition vector of the 6d $\cN=(2,0)$ theory on a closed 6d manifold \emph{together with codimension-two defects}.  Currently, the properties of the codimension-two defects are inferred by studying its behavior in lower dimensional compactifications, and it seems difficult to directly study the behavior of the partition vector in 6d.  Using the holographic dual of the codimension-two defects found in \cite{Gaiotto:2009gz} might be useful; after all, the behavior of the partition vector of the 6d $\cN=(2,0)$ theory was first found in this holographic context \cite{Aharony:1998qu,Witten:1998wy}.
\item To extend the analysis in this note to the spacetimes with torsion cycles and/or non-Spin manifolds. Incorporating the case with torsion cycles will be important to study the lens space index, i.e.~the partition function on $S^3/\bZ_r\times S^1$ \cite{Benini:2011nc,Razamat:2013opa}. 
Some supersymmetric theories can be formulated on non-Spin but Spin$_C$ manifolds. In the case of 6d $\cN=(2,0)$ theory of general type $\fg$, there can  be more possibilities. These points would be important to study class $S$ theories on $\mathbb{CP}^2$, for example. 
\item To study the behavior of external line operators of class $S$ theories in more detail. They arise from external codimension-4 operators of the 6d $\cN{=}(2,0)$ theory, and can be analyzed from this point of view. This will tell us not only the discrete charges of the line operators, but also a more detailed structure, corresponding to the distinction of the center $\cC=\{\text{weight lattice}\}/\{\text{root lattice}\}$ and the weight lattice itself. 
\end{itemize} 
 The author hopes to come back to some of the issues listed above  in the future.
 
\section*{Acknowledgements}
It is a pleasure for the author to thank    A. Neitzke, K. Ohmori, S. Razamat, N. Seiberg, N. Watanabe and B. Willet for helpful discussions. He would also like to thank O. Aharony,  G. W. Moore and N. Seiberg in particular for carefully reading the manuscript and giving many illuminating comments, and J. Distler for pointing out an error in the v1 of the preprint.
The author is  supported in part by JSPS Grant-in-Aid for Scientific Research No. 25870159,
and in part by WPI Initiative, MEXT, Japan at IPMU, the University of Tokyo.
The work was completed during the author's stay at the Aspen Center for Physics  
and at the Simons Center for Geometry and Physics.
The author thanks these two institutions for their generous hospitality, and was
partially supported by the NSF Grant \#1066293 
 during his visit at the Aspen Center.

\bibliographystyle{ytphys}
\baselineskip=.92\baselineskip
\let\bbb\bibitem\def\bibitem{\itemsep1pt\bbb}
\bibliography{ref}
\end{document}